# Agent-Based Modeling for Multimodal Transportation of $CO_2$ for Carbon Capture, Utilization, and Storage: CCUS-Agent[†]

**Majbah Uddin, PhD (*Corresponding Author*)[‡]**
Oak Ridge National Laboratory
1 Bethel Valley Road, Oak Ridge, TN 37830
Email: uddinm@ornl.gov

**Robin J Clark, MS[‡]**
Oak Ridge National Laboratory
1 Bethel Valley Road, Oak Ridge, TN 37830
Email: clarkrj@ornl.gov

**Michael R Hilliard, PhD**
Oak Ridge National Laboratory
1 Bethel Valley Road, Oak Ridge, TN 37830
Email: hilliardmr@gmail.com

**Joshua A Thompson, PhD**
Oak Ridge National Laboratory
1 Bethel Valley Road, Oak Ridge, TN 37830
Email: joshua.thompson@chevron.com

**Matthew H Langholtz, PhD**
Oak Ridge National Laboratory
1 Bethel Valley Road, Oak Ridge, TN 37830
Email: langholtzmh@ornl.gov

**Erin G Webb, PhD**
Oak Ridge National Laboratory
1 Bethel Valley Road, Oak Ridge, TN 37830
Email: webbeg@ornl.gov

[†] This manuscript has been authored by UT-Battelle, LLC, under contract DE-AC05-00OR22725 with the US Department of Energy (DOE). The US government retains and the publisher, by accepting the article for publication, acknowledges that the US government retains a nonexclusive, paid-up, irrevocable, worldwide license to publish or reproduce the published form of this manuscript, or allow others to do so, for US government purposes. DOE will provide public access to these results of federally sponsored research in accordance with the DOE Public Access Plan (http://energy.gov/downloads/doe-public-access-plan).
[‡] Contributed equally.

**Abstract**

To understand the system-level interactions between the entities in Carbon Capture, Utilization, and Storage (CCUS), an agent-based foundational modeling tool, CCUS-Agent, is developed for a large-scale study of transportation flows and infrastructure in the United States. Key features of the tool include (i) modular design, (ii) multiple transportation modes, (iii) capabilities for extension, and (iv) testing against various system components and networks of small and large sizes. Five matching algorithms for $CO_2$ supply agents (e.g., powerplants and industrial facilities) and demand agents (e.g., storage and utilization sites) are explored: Most Profitable First Year (MPFY), Most Profitable All Years (MPAY), Shortest Total Distance First Year (SDFY), Shortest Total Distance All Years (SDAY), and Shortest distance to long-haul transport All Years (ACAY). Before matching, the supply agent, demand agent, and route must be available, and the connection must be profitable. A profitable connection means the supply agent portion of revenue from the 45Q tax credit must cover the supply agent costs and all transportation costs, while the demand agent revenue portion must cover all demand agent costs. A case study employing over 5,500 supply and demand agents and multimodal CCUS transportation infrastructure in the contiguous United States is conducted. The results suggest that it is possible to capture over 9 billion tonnes (GT) of $CO_2$ from 2025 to 2043, which will increase significantly to 22 GT if the capture costs are reduced by 40%. The MPFY and SDFY algorithms capture more $CO_2$ earlier in the time horizon, while the MPAY and SDAY algorithms capture more later in the time horizon.

**Keywords**
Carbon capture, utilization, storage; Agent-based model; Multimodal $CO_2$ transportation; 45Q tax credit

## 1. Introduction

Carbon capture, utilization, and storage (CCUS) technologies are essential to most deep decarbonization pathways. It includes decarbonizing industry and powerplants by capturing emissions and $CO_2$ directly from the air and bioenergy with carbon capture and sequestration. The captured $CO_2$ is either utilized in another industry, converted to fuels or products (e.g., building materials and plastics), stored for later use, or sequestered and permanently removed from the atmosphere via underground storage or in a solid form. To reach a net zero goal by 2050 in the United States, the integration of CCUS and the related transportation infrastructure involving multiple modes (i.e., pipeline, water, rail, and truck) must be accomplished at a multi-gigaton scale (Fahs et al., 2023; NASEM, 2023). Incentives such as the 45Q tax credit provide the basis for a carbon economy that can decrease emissions or even offer a negative carbon cycle. CCUS project developers can use this credit as a reliable and performance-based income stream. The Inflation Reduction Act of 2022 increased the tax credit amount and decreased the threshold for CCUS projects, providing further support to the CCUS system (IRS, 2023). For example, captured $CO_2$ from industries and powerplants to be stored in saline and other geologic formations is now eligible for a $85 per tonne credit once the facility is placed in service, a 70% increase from $50 per tonne. The annual $CO_2$ capture threshold to receive the tax credit for powerplants was reduced to 18,750 tonnes from 500,000 tonnes. To fully support the planning and development of CCUS transportation infrastructure, there is a need for a modeling framework that can incorporate multi-modal transportation options on a large scale, CCUS system evolution, and complex interactions among the entities in the system.

Most of the current models in the CCUS supply chain and transportation arena are focused on optimal matching of sources and sequestration points and optimizing pipeline infrastructures (Ahn et al., 2020; Alhajaj & Shah, 2020; Becattini et al., 2022; Hasan et al., 2014, 2015, 2022; Leonzio et al., 2020; Ma et al., 2023; Middleton et al., 2020; Ostovari et al., 2022; Tapia et al., 2018; Wu et al., 2022; Zhang et al., 2018, 2020). The models use classic optimization approaches leveraging graph theory and integer programming techniques. Hasan et al. (2014) developed a supply chain network optimization model for CCUS that minimized costs to reduce $CO_2$ emissions from point sources and their negative environmental impacts. The model was applied to nationwide, regional, and statewide systems, and it was found that 50 to 80% of the current $CO_2$ emissions could be reduced. In another work, Hasan et al. (2015) presented a hierarchical and multi-scale framework for CCUS and carbon capture and utilization supply chain networks, considering minimum capital, operational, and materials costs. The framework is applied to the U.S. national network, and it was mentioned that it could also be applied to regions and other countries. Tapia et al. (2018) presented an in-depth discussion of optimization and decision models for the planning of CCUS. They recommended focus be placed on developing advanced tools that can be used for large-scale implementation of CCUS technology. Zhang et al. (2018) developed a mixed integer linear programming model for optimizing the CCUS supply chain in Northeastern China. The model can select $CO_2$ sources, capture, utilization, storage facilities, and pipeline networks as intermediate transportation facilities simultaneously. To give the geographic distribution and magnitude of capture, use, and sequestration sites, as well as the transportation routes for various scenarios, the entire network has been economically optimized for over 20 years. Zhang et al. (2020) extended the previous model to design a more environmentally friendly CCUS supply chain via multi-objective optimization modeling.

Ahn et al. (2020) developed a mixed integer linear programming model for enhanced shale gas recovery by $CO_2$ storage (via injection). Leonzio et al. (2020) analyzed the CCUS supply chain in Italy and Germany using mixed integer linear programming models. Unique features of the modeling include the consideration of carbon tax remission, economic incentives, and revenues subtracted from the total supply chain cost. Alhajaj and Shah (2020) developed a multiscale optimization model for carbon capture and storage supply chain, and the model was applied to the United Arab Emirates case with capture plant, rail, and pipelines for $CO_2$. Middleton et al. (2020) developed an open-source tool called SimCCS$^{2.0}$ that can formulate an optimization model to determine the cost-optimized carbon capture and storage system design as well as generate candidate pipeline transportation routes based on custom inputs on $CO_2$ source, sink, and data related to the transportation network. Becattini et al. (2022) presented an optimization framework for the CCUS supply chain considering the construction, scaling, and operation of $CO_2$ capture and transport technologies. The model minimizes the total cost while considering different emissions reduction pathways and deployment of technologies over 25 years. Ostovari et al. (2022) developed a climate-optimal supply chain for CCUS through mineralization to quantify the large-scale mineralization potential in Europe. Wu et al. (2022) developed an optimization model to match sources and sinks in the CCUS system. The model can incorporate dynamics of capacity expansion associated with CCUS activities. More recently, Ma et al. (2023) developed SimCCS$^{3.0}$, an extension of SimCCS$^{2.0}$, to include new features, including multiple periods for carbon capture and storage project life and dynamic evolution of source status. The tool is tested for two optimization case studies involving regional systems.

The optimization approach, however, assumes a centralized, perfectly informed, rational design. Because the CCUS system can be viewed as a market with agents (supply and demand) who respond to (tax) incentives, alternate methods like agent-based simulation modeling could better model the interactions between the agents. Agent-based modeling has several benefits: (1) it can capture emergent phenomena, (2) it can provide a natural description of a system, and (3) it is flexible, i.e., new agents can be added easily to the system (Bonabeau, 2002). Another benefit of designing and implementing an agent-based model is the identification of key assumptions and parameters for the market system.

To this end, to understand the system-level interactions among the agents in the CCUS system, a foundational modeling tool called *CCUS-Agent* for a large-scale (i.e., United States) study of CCUS transportation flows and infrastructure is developed. Based on the authors' best knowledge, the CCUS-Agent is the first tool to incorporate CCUS transportation flows on multimodal networks and associated infrastructure on a large scale and tax incentives for CCUS. The key features of the tool are summarized as follows:
- **Modular design.** The modeling architecture is flexible and extendable to model the complexity and size of the entire system at an appropriate level of detail. One advantage of this design is that the user only needs to update the relevant module when there are changes to the agents in terms of infrastructure and other plans in the future.
- **Multiple transportation modes.** The tool can use waterway, rail, truck, and pipeline movements, considering the system's evolution, distances, locations, and volumes. Additionally, the tool can consider rolling buildout of transportation infrastructure (e.g., pipeline networks).
- **Extendable.** The tool can be extended to select locations for some future capture facilities based on facility costs, transportation, and sequestration options.
- **Verified.** The tool is tested against various system components and networks of small and large sizes.

The paper contributes to the literature in several ways. First, it develops a modeling framework to support CCUS transportation infrastructure planning with multiple types of $CO_2$ sources and sinks, and waterway, rail, truck, and pipeline movements on large-scale transportation networks. Second, it explicitly considers external support, such as tax incentives, in CCUS to model the system's evolution more realistically for achieving a net-zero energy system by 2050. Lastly, it explores different types of costs relevant to CCUS operations and quantifies the increase in total system $CO_2$ capture with the reduction of those costs. Thus, the paper provides information and tools for policy-makers and stakeholders of CCUS that can guide better planning of transportation infrastructure and flows and better address the challenges of climate change.

## 2. Methods

### 2.1. Problem Statement

The CCUS-Agent is designed to maximize $CO_2$ capture while determining the potential transportation constraints the CCUS system might encounter over time. The following assumptions are made for the addressed problem:

- The purpose of the supply agents (e.g., powerplants and industrial facilities) is to capture $CO_2$. Each supply agent has a property called agent type. The routing path for some agent types is limited. For example, the food-grade uses of $CO_2$ are prohibited from using pipelines.
- The demand agents (e.g., storage and utilization sites) sequester or use $CO_2$. Each demand agent has a property called agent type.
- Each agent has an adjustable start date property; this allows one to emulate that not everyone will adopt at the beginning of time. All supply agents' start dates are randomly set between 2025 and 2032. This is based on the 45Q tax credit provision that requires that qualified projects commence construction by the end of 2032 (IRS, 2023). The supply agent adoption rate could be influenced by profit, the distance to the transportation network, or $CO_2$ volume.
- The storage demand agents have unlimited capacity (Fahs et al., 2023). The current sites have been carefully placed across the region. If too much volume goes to a single location, then this would indicate that another site might need to be developed in the same area.
- Revenue is provided to the demand agents based on the agent's $CO_2$ volume and type. The demand agents share a percentage of the revenue with the supply agents. The demand agent's portion of the revenue must cover their cost only. The supply agent portion of the revenue must cover not only their cost but also all relevant transportation capital investment and operating costs.

### 2.2. Model and Agent Flow

The CCUS-Agent model is created so that the supply and demand nodes act as agents in the system. Figure 1 shows the agent flow in the model; their current state drives it. Both supply and demand agents are held upstream at their respective entrances and are released into the system when the simulation clock matches their start date. The agents then transition into the selection state. While the supply agents remain in the selection state, they evaluate their potential connections annually as the transportation routes evolve during the simulation. When supply agents find a connection, they transition from the selection state into a connected state. Supply agents remain connected for 12 years based on the 45Q tax credit provision (IRS, 2023). However, the duration has been parameterized and can be adjusted. Once the connection duration has been completed, the supply agents will transition to a complete state and exit the system. Each demand agent has a property of annual capacity, total capacity, and end date. However, for this study, demand agents are assumed to have an infinite capacity, and they will remain available for any supply agent to connect to for the duration of the simulation run.

*2.3. Transportation Routes*

Transportation routes are assumed to have three segments: A. first leg (i.e., routes connecting supply agents to the nearest nodes in the network), B. middle leg (i.e., long-haul transportation routes), and C. last leg (i.e., route connecting demand agents to the nearest nodes in the network). For the water mode, trucks are used for transportation over Legs A and C with intermodal terminals at the ends to help transfer $CO_2$ from one mode to another. Spur lines are added to and from the supply and demand facilities for rail. For pipelines, private pipelines are added to and from the facilities. The capital investment costs for the construction of Legs A and C for pipeline and rail, as well as terminals for water, are included in the total system cost. The model uses the existing rail and water network for Leg B. It is assumed that Leg B for the $CO_2$ pipeline would evolve continuously with external support (e.g., $CO_2$ transport FEED project (USDOE, 2023)). However, the agents know when the routes will come online and can choose to postpone their connection to a demand agent. The rail and water routes are available at the beginning of the simulation and for the entirety of the time horizon.

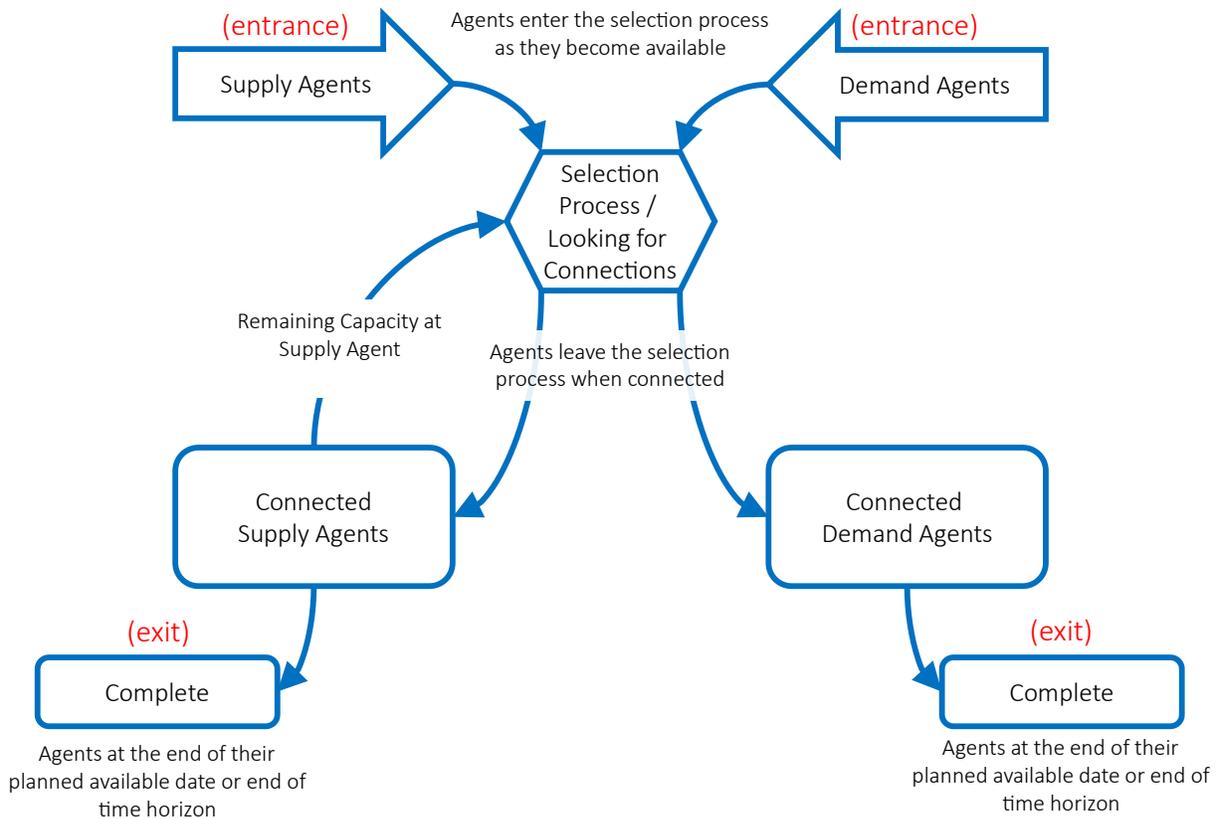

**Figure 1. Supply and demand agent flow in the model.**

*2.4. Source-Sink Matching*

Reasonable source-sink matching planning could substantially reduce the CCUS system overhead (Patricio et al., 2017; Wu et al., 2022). CCUS-Agent contains five selection algorithms for source-sink matching, i.e., the mechanism for the supply agents to select both a demand agent and a specific route to the demand agent. In all algorithms, before a connection can commence, the supply agent, demand agent, and route must currently be available (the simulation date must be equal to or greater than the available date), and the connection must be profitable. A profitable connection means the supply agent portion of revenue must cover the supply agent costs and all transportation costs. In contrast, the demand agent revenue portion must only cover all demand agent costs. Some algorithms prefer to connect agents during the first year when there is a profitable connection, while other algorithms consider the future pipeline and demand agent development plans and delay their connection until a better route or demand agent becomes available. The algorithms are selected considering two main criteria: profit and distance. For profit criterion, the agents evaluate the first profitable connection or wait and look ahead to choose the most profitable connection. For distance criterion, the matching is tested for shortest total distance in the first year, shortest total distance in all years, or shortest distance that minimizes capital investment costs for private infrastructure. Note that the comparison for matching in the first year against the best year in the planning horizon helps to assess the impact on the transportation mode choice (e.g., pipeline vs. rail) and associated costs. The algorithms and a brief description are provided below:

1. **Most Profitable First Year (MPFY).** Select the most profitable connection within the first profitable year.
2. **Most Profitable All Years (MPAY).** Select the most profitable connection while considering future transportation infrastructure development plans. The connection will not begin until the route is available.
3. **Shortest Total Distance First Year (SDFY).** Select the shortest total distance connection within the first profitable year. Only profitable connections are considered.
4. **Shortest Total Distance All Years (SDAY).** Select the shortest total distance connection while considering future transportation infrastructure development plans. The connection will not begin until the route is available. Only profitable connections are considered.
5. **Shortest distance to long-haul transport All Years (ACAY).** Select the shortest first and last leg distance connection while considering future transportation infrastructure. This objective focuses on minimizing capital investment costs for private infrastructure. The connection will not begin until the route is available. Only profitable connections are considered.

*2.5. Model Interface*

The model is built in ExtendSim v10, taking full advantage of the internal program database by integrating it into all model constructs. The model database contains all routing paths and agent properties, including their inputs, outputs, and current agent state. Figure 2 shows the CCUS-Agent main worksheet, and Figure 3 shows a screenshot of the model database.

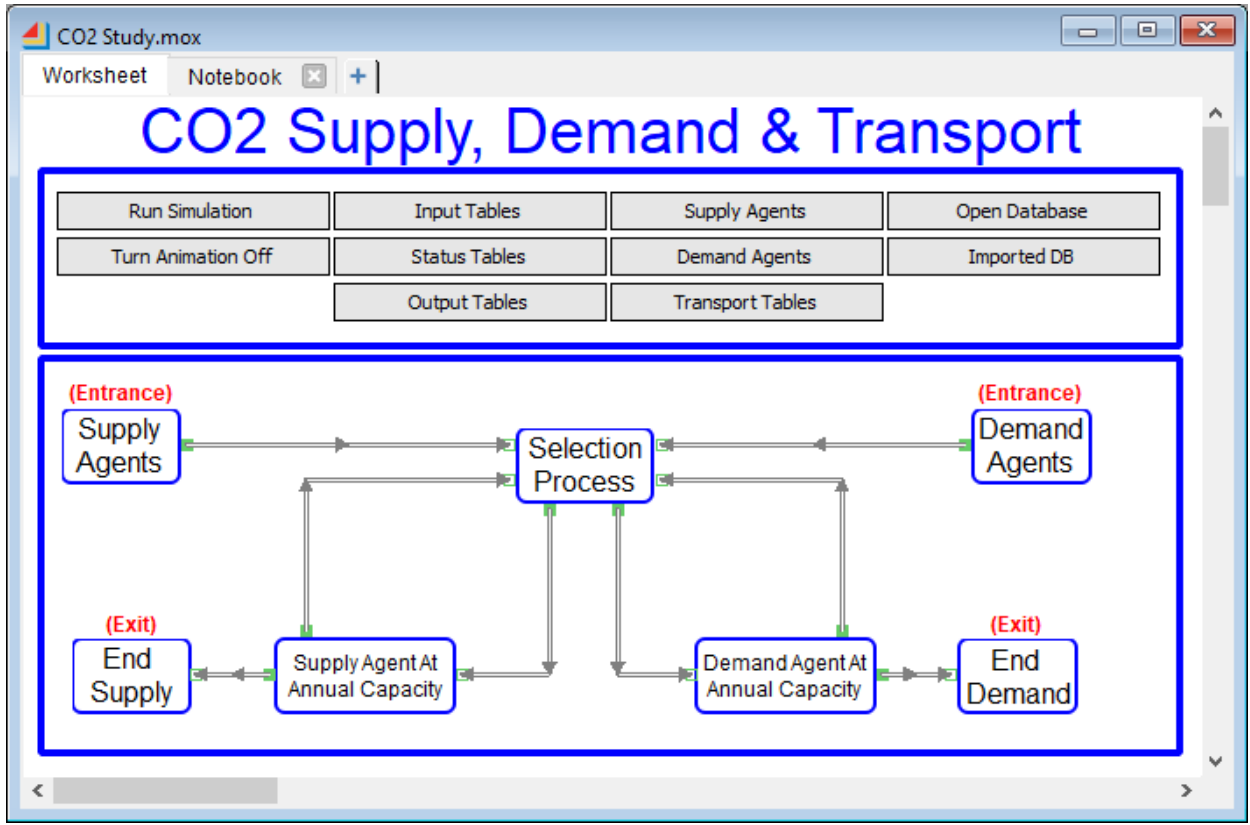

**Figure 2. CCUS-Agent main worksheet.**

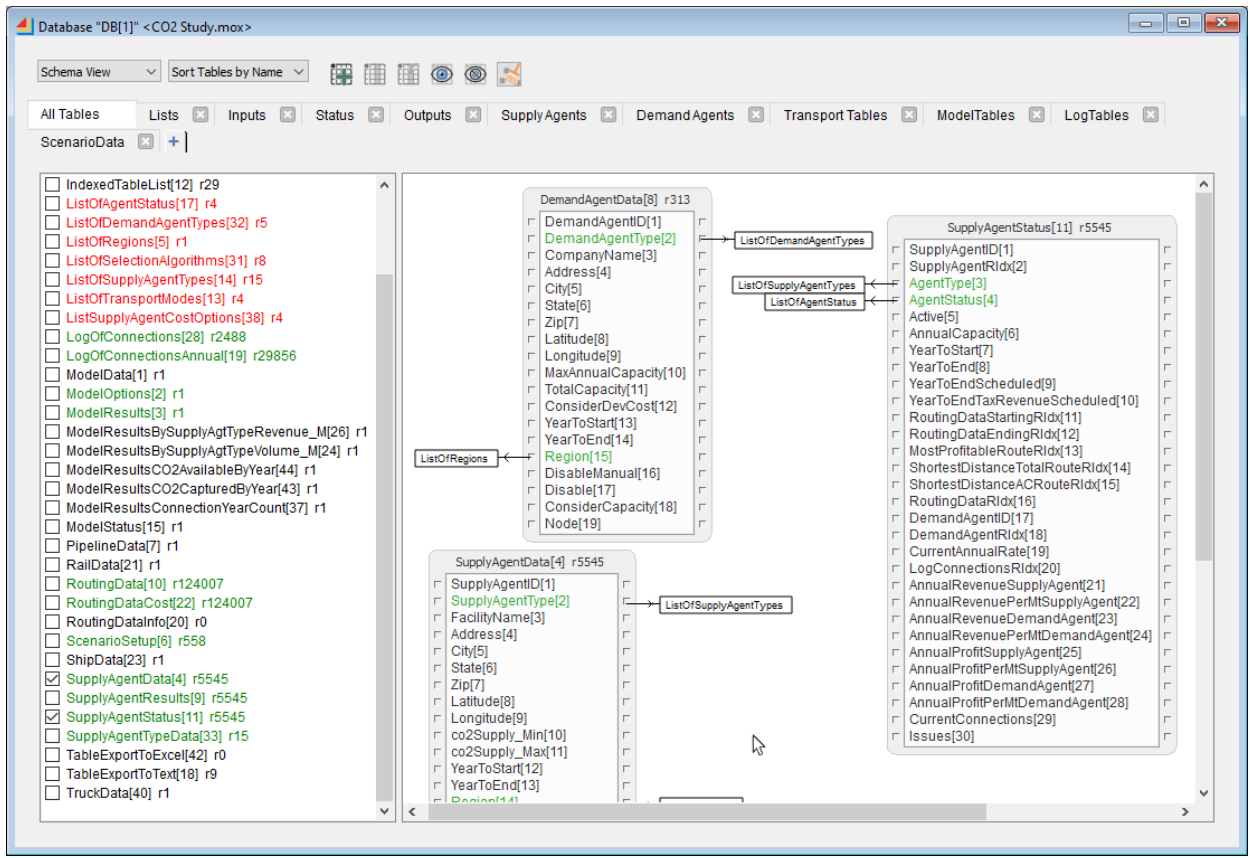

**Figure 3. Snapshot of CCUS-Agent database.**

## 3. Data Preparation

The following subsections describe data used for a case study to test CCUS-Agent for $CO_2$ sources and sinks, distances over multimodal transportation networks, agent costs (i.e., capture and storage costs), transportation costs, and revenue from the 45Q tax credit.

### 3.1. Sources

Significant greenhouse gas (GHG) emission sources, fuel and industrial gas suppliers, and $CO_2$ injection sites in the United States must report annual data about GHG emissions to the Environmental Protection Agency (EPA). These data are available from the Facility Level Information on Greenhouse Gases Tool (EPA, 2020). Table 1 categorizes all these sources into 13 groups: Powerplants, Petroleum and natural gas systems, Refineries, Chemicals, Cement, Iron and Steel, Minerals, Petrochemicals, Pulp and paper, Hydrogen, Waste, Metals, and Others. Associated emissions in 2020 from the sources are also provided. Additionally, candidate Bioenergy with Carbon Capture and Storage (BECCS) plants and associated $CO_2$ quantities are collected from the U.S. Department of Energy (2020). Lastly, candidate stationary direct air capture (DAC) locations are considered based on Abramson et al. (2023). Among the 5,544 sources, the most significant is the powerplants, with approximately 1,500 million tonnes per annum (MTPA), over 52% of the $CO_2$ supply amount.

**Table 1. Types of stationary sources and their $CO_2$ quantity.**

| Source Type | Number of Sources | $CO_2$ Quantity (MTPA) | Share of $CO_2$ Supply |
|---|---:|---:|---:|
| Powerplant | 1,158 | 1,476.8 | 52.1% |
| Bioenergy with Carbon Capture and Storage | 80 | 502.7 | 17.7% |
| Petro & NG Systems | 1,804 | 231.8 | 8.2% |
| Refinery | 133 | 181.6 | 6.4% |
| Chemicals | 275 | 91.0 | 3.2% |
| Other | 1,043 | 76.2 | 2.7% |
| Cement | 91 | 66.2 | 2.3% |
| Iron & Steel | 119 | 62.3 | 2.2% |
| Minerals | 273 | 42.6 | 1.5% |
| Petrochemicals | 56 | 40.2 | 1.4% |
| Pulp & Paper | 218 | 34.4 | 1.2% |
| Hydrogen | 37 | 15.5 | 0.5% |
| Waste | 79 | 14.0 | 0.5% |
| Metals | 154 | 13.1 | 0.5% |
| Direct Air Capture | 24 | 6.3 | 0.2% |
| **All** | **5,544** | **2,854.6** | **100.0%** |

Figure 4 illustrates the spatial distribution of the $CO_2$ sources along with their proximity to multimodal transportation networks. As evident from the size of the circle, the $CO_2$ sources are scattered in different regions with varying supply amounts. Part *a* of the figure shows sources with $CO_2$ trunk and spur pipelines in 2050 collected from Princeton's Net-Zero America study (NZA, 2023a, 2023b). The pipeline geospatial data is provided from 2020 to 2050, with network updates every 5-year interval. Part *b* of the figure shows sources with North American Rail Network Lines collected from the U.S. Department of Transportation (USDOT, 2024a). The network lines are real-world rail transportation tracks. Lastly, part *c* of the figure shows sources with Navigable Waterway Network Lines collected from the U.S. Department of Transportation (USDOT, 2024b).

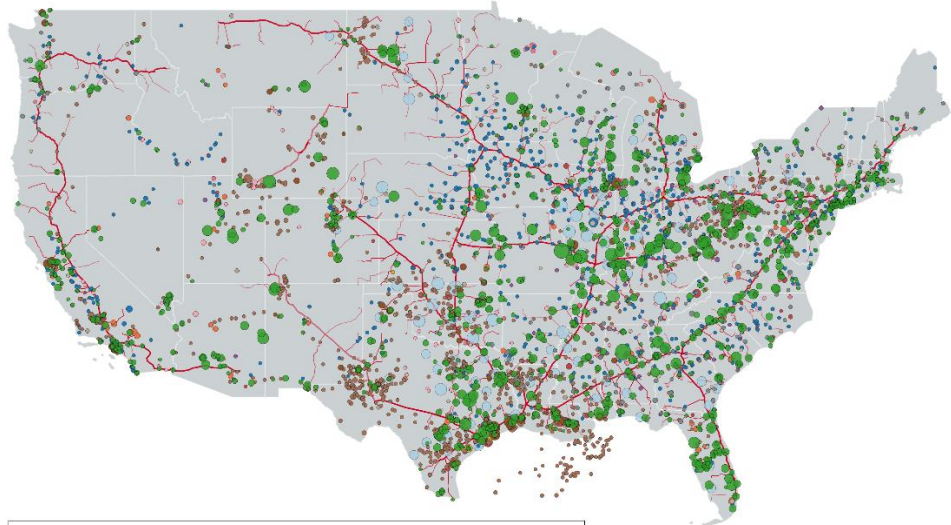
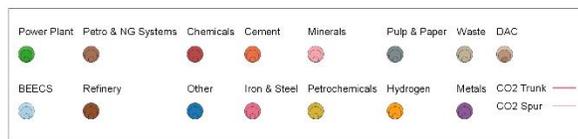
(a)
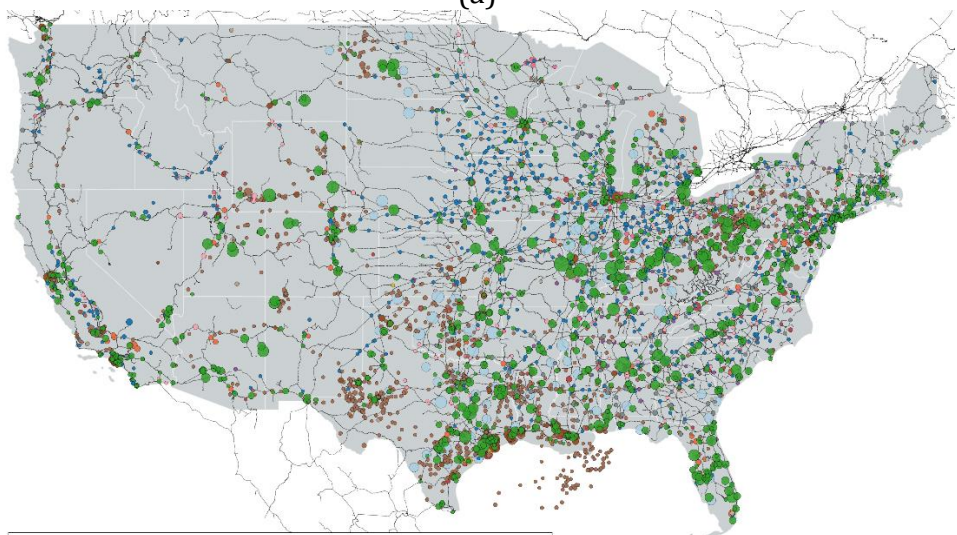
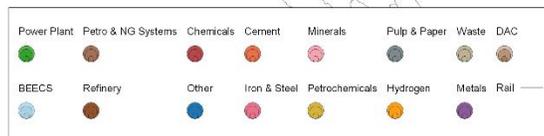
(b)

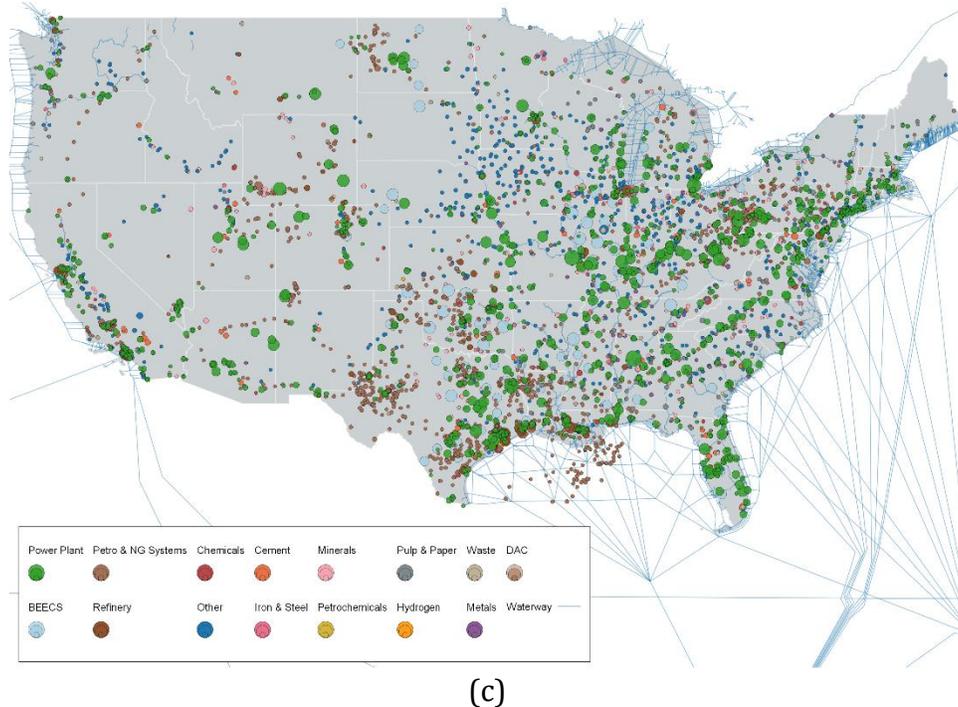

(c)

**Figure 4. $CO_2$ sources with transportation networks: a) Pipeline, b) Rail, and c) Water.**

*3.2. Sinks*

This study considers two major types of $CO_2$ sinks: storage sites and utilization facilities. There are only a few operating geologic storage sites in the United States (Fahs et al., 2023). For better geographical representation of the storage sites, a multi-criteria decision approach is applied to identify candidate storage sites in this study. Figure 5 presents the details of the methodology. The input data are GHG facility (i.e., source sites), transportation networks, population count (Weber et al., 2020), and assessment unit regions (USGS, 2013, 2022). First, facility clusters were generated at every 100-mile distance, and random sites were generated in the regions as potential storage sites. Then, a total of 120 candidate storage sites were selected based on three criteria: (1) the sites are near the transportation network, (2) near facility clusters, and (3) located in low-risk areas (i.e., low population count during daytime and nighttime). This methodology for candidate storage site selection is implemented in geospatial software QGIS. A custom script is written that takes input all the required datasets and then applies a filter procedure considering the three criteria mentioned above to generate the outputs for the candidate storage sites. Figure A.1 in the appendix shows these selected storage sites along with all input data layers. The storage sites are scattered across the United States.

For utilization, urea production plants, food and beverage plants, enhanced oil recovery locations, and other use locations are considered. Urea synthesis is currently the most significant volume of $CO_2$ utilization (NASEM, 2023). The location of urea plants and the demand for $CO_2$ are compiled from Statisa (2019) and Knoema (2023). The locations for food and beverage industries are identified using North American Industry Classification

System (NAICS) code 3121 (Beverage manufacturing) and 31141 (Frozen food manufacturing) from the County Business Pattern dataset (US Census Bureau, 2024). Enhanced Oil Recovery (EOR) locations are compiled from Advanced Resources International (2021). Lastly, the locations for other utilization are identified based on carbon capture and utilization (CCU) startups from CCU Activity Hub (Global $CO_2$ Initiative, 2023). Figure 6 illustrates the distribution of the $CO_2$ sinks along with their proximity to pipeline (part *a*), rail (part *b*), and water (part *c*) transportation networks.

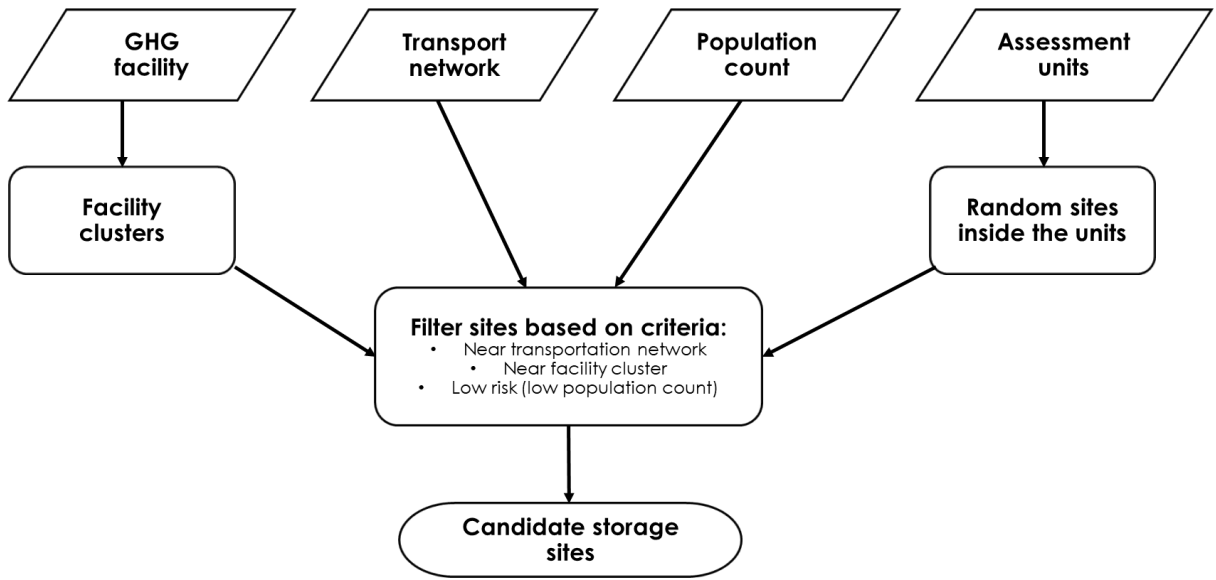

**Figure 5. Methodology for identifying candidate storage sites.**

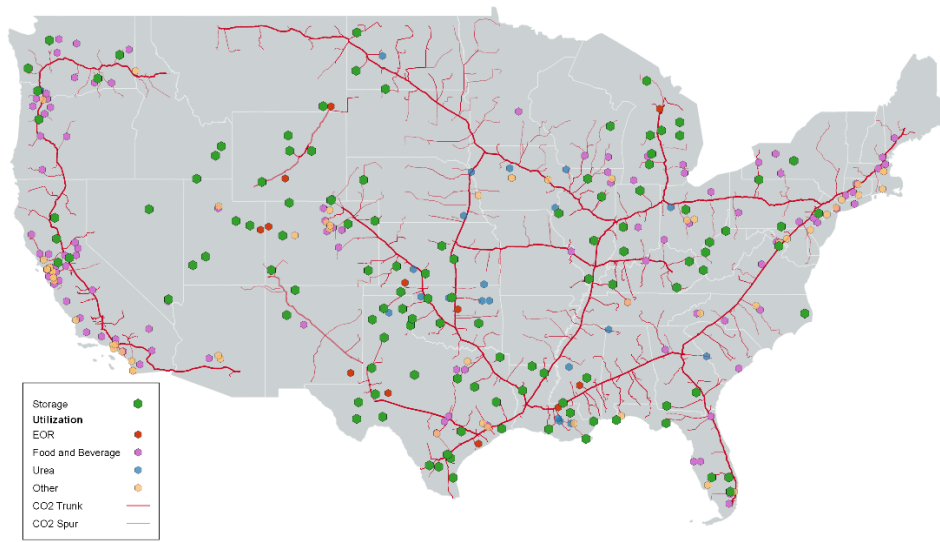

(a)

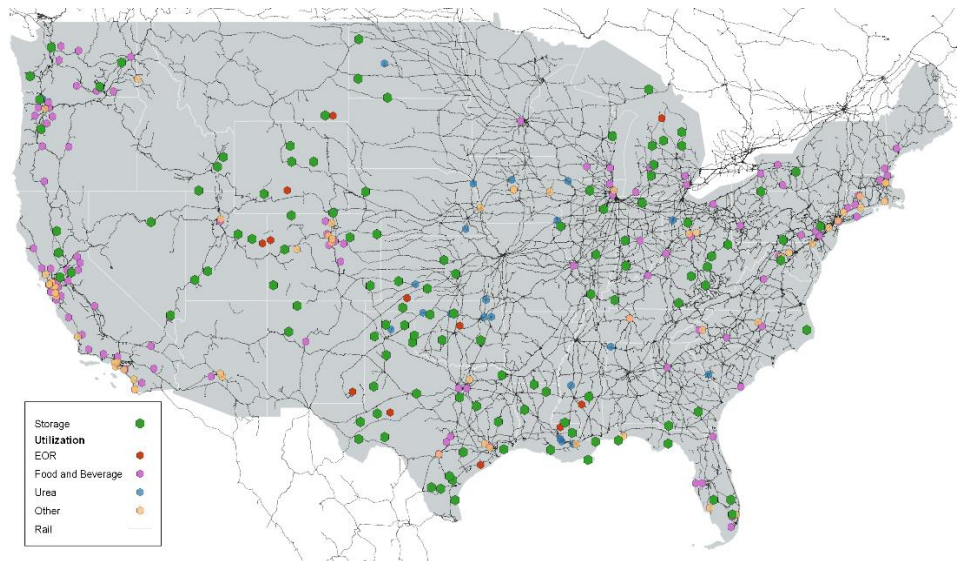

(b)

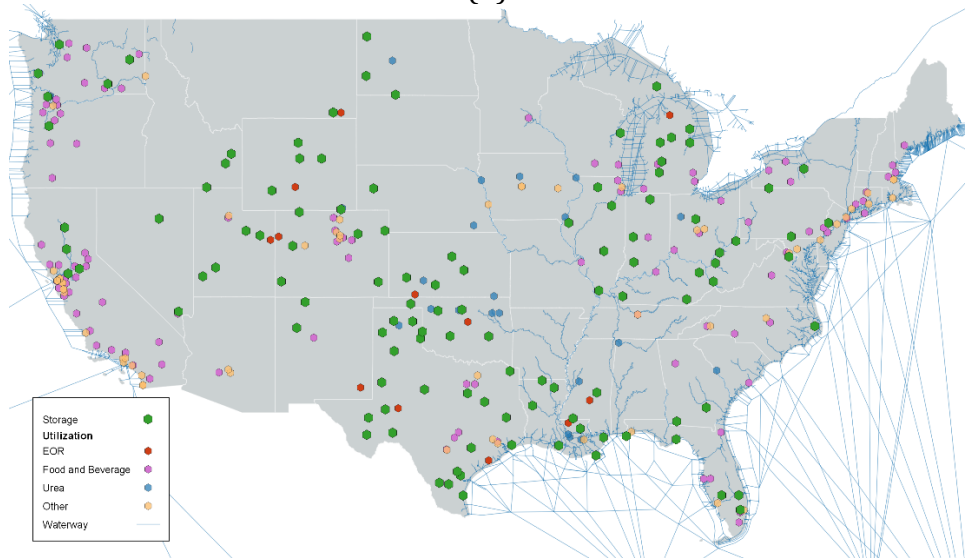

(c)

**Figure 6. CO$_2$ sinks with transportation networks: a) Pipeline, b) Rail, and c) Water.**

*3.3. Network Distances*

For rail and water transportation modes, the first segment (leg A) and last segment (leg C) distances are calculated based on great circle distances between the source and the nearest node in the network and the sink and the nearest node, respectively. The middle segment (leg B) distances are calculated using the network's shortest path lengths. The pipeline's three segment distances are calculated based on the great circle distance between the respective nodes. At the end of this calculation step, three distance matrices are finalized, one for each transportation mode with a dimension of source x sink.

*3.4. Agent and Transportation Costs and Revenue*

Capture costs for supply agents by type are collected from Baylin-Stern and Berghout (2023), as shown in Table 2. They provide low and high estimates for capture costs. It is assumed that the value would follow a Uniform distribution within the range. This helps with the simulation testing for multiple replications of the model instances by randomly drawing a value from the distribution. Storage cost (storage in underground geologic formations) is 10 USD per tonne of $CO_2$ stored (Baylin-Stern & Berghout, 2023). Table 3 presents the revenue from the 45Q tax credit (IRS, 2023) by source and sink type. Lastly, Table 4 presents the transportation costs by mode and capital investment costs for pipeline and rail network and intermodal terminal to be used in water transportation. The references for these data are also provided in the table.

**Table 2. Capture cost input for supply agents.**

| Source Type | USD per tonne |
|---|---|
| DAC | U[134, 342] |
| BECCS | U[55, 60] |
| Cement | U[60, 120] |
| Chemicals | U[15, 25] |
| Hydrogen | U[50, 80] |
| Iron & Steel | U[40, 100] |
| Metals | U[40, 100] |
| Minerals | U[15, 25] |
| Other | U[15, 25] |
| Petro & NG Systems | U[15, 25] |
| Petrochemicals | U[15, 25] |
| Power Plant | U[50, 100] |
| Pulp & Paper | U[40, 100] |
| Refinery | U[15, 25] |
| Waste | U[40, 100] |

U represents Uniform distribution

**Table 3. Revenue.**

| Sink Agent Type | Source Agent Type | Revenue (USD per tonne) |
|---|---|---|
| Storage (Underground Geologic Formations) | Industrial and power generation facilities | 85 |
| Storage (Underground Geologic Formations) | DAC | 180 |
| Utilization | Industrial and power generation facilities | 60 |
| Utilization | DAC | 130 |

**Table 4. Transportation-related cost input.**

| Mode | Transportation Cost | | | Capital Investment Cost | | |
| --- | --- | --- | --- | --- | --- | --- |
| | Value | Unit | Source | Value | Unit | Source |
| Pipeline | 0.0161 | USD per tonne-mile | Stolaroff et al. (2021) | 784,198.0 | USD per mile | Greig et al. (2021) |
| Rail | 0.0708 | USD per tonne-mile | Stolaroff et al. (2021) | 2,000,000.0 | USD per mile | NASEM (2015) |
| Water | 0.0644 | USD per tonne-mile | Roussanaly et al. (2021) | 4,585.1 | USD per tonne | Dominion Energy (2015) |
| Truck | 0.1770 | USD per tonne-mile | Stolaroff et al. (2021) | | | |

## 4. Results and Discussion

Using the data presented above, several numerical experiments were run in CCUS-Agent. Note that profitable source-sink connections are only considered—a connection where the revenue from the tax credit would cover the supply and demand agent costs and the transport costs for the duration of the connection. The system starts in 2025 and allows new agent connections up to 2032; once connected, the source and sink agents will receive a tax credit for 12 years. Based on the above, the time horizon is from 2025 to 2043. The supply agents are assumed to have $CO_2$ available annually, as presented in Table 1.

The first experiment compares the different source-sink matching algorithms. The key difference between the scenarios in the experiment is the selection algorithm used, which affects the chosen path. The run-to-run variability within a scenario is due to several random parameters: the starting dates assigned to the agents, the percentage of $CO_2$ captured, and the capture cost of each supply agent type. Because the system is unconstrained, the same feasible connections will be available in all algorithms but with different starting dates; therefore, the total $CO_2$ captured should be similar for all algorithms. Figure 7 shows the total $CO_2$ captured in billion tonnes (GT) for the algorithms. The distribution of the captured amount is from 30 replications of the experiment—the median total $CO_2$ captured over the period 2025 to 2043 ranges from 9.12 to 9.81 GT.

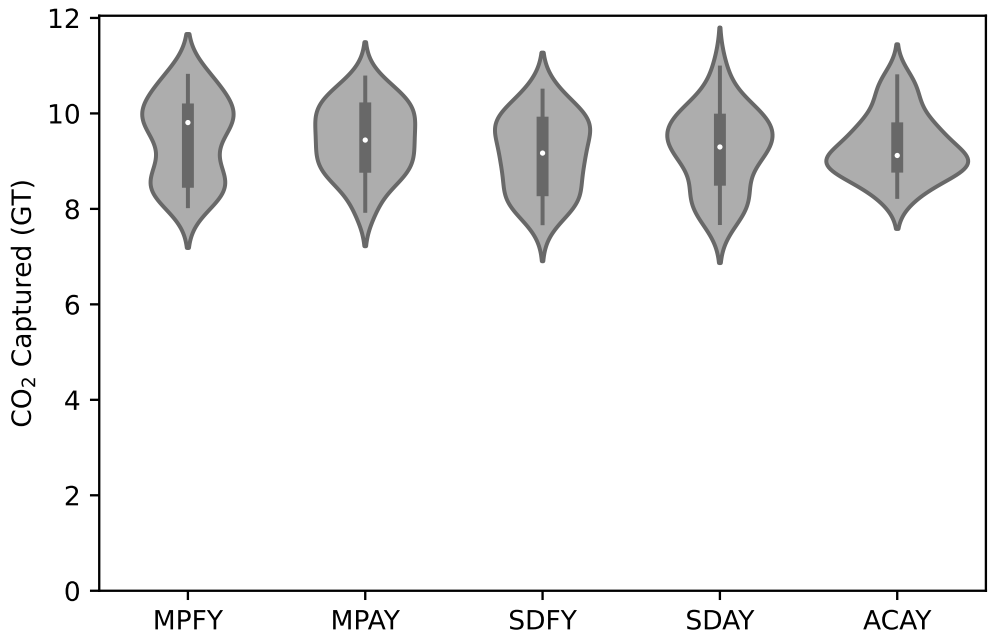

**Figure 7. Total $CO_2$ captured from 2025 to 2043 by matching algorithm.**

Figure 8 displays the share of $CO_2$ captured and transported by transportation mode for the five algorithms. Rail and water transport are the primary modes under the MPFY, SDFY, and SDAY algorithms. The MPFY and SDFY algorithms select the first profitable route, which indicates that the rail and water routes tend to be the first most profitable early in the time horizon. The SDAY algorithm selects the shortest distance, which means the profitable rail and water modes would have the shortest distance between agents. If MPFY and MPAY are compared, one can see that rail and water modes are preferred when choosing the first profitable route; however, the pipeline mode replaces a sizeable portion of the water mode when considering future routes. Comparing the shortest total distance using SDFY and SDAY, there was no significant difference in transportation modes, nor were there any significant differences compared to the MPFY algorithm. The rail mode dominates the ACAY algorithm, with roughly equal water and pipeline mode shares. By comparing MPAY and ACAY, which look across all years, it can be inferred that even though some rail routes might require less new development, they would be more expensive than pipelines and water. Note that pipelines are not the dominant transportation mode due to the unavailability of a large-scale network for most of the time horizon and longer distances to long-haul transportation networks.

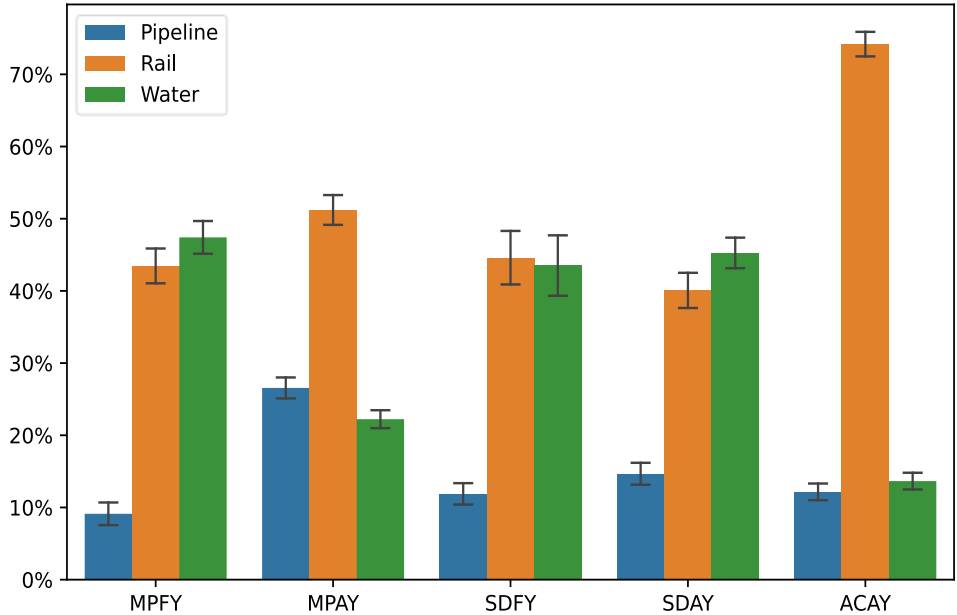

**Figure 8. Share of $CO_2$ by transportation mode and matching algorithm.**

A summary of distances between connected agents is presented in Table 5. There is a significant difference in the total distance between connected agents comparing the MPFY and MPAY algorithms. The difference between these two algorithms shows that the transportation distance can be reduced by over 50% when considering future pipeline development and waiting for the most suitable time to set up a new connection. However, just minimizing the first and last leg distance with the ACAY algorithm almost doubles the total distance compared to the SDAY algorithm. The MPAY, SDFY, and SDAY algorithms have similar distances, indicating that selecting the most profitable connection route while looking ahead will minimize the total distance.

**Table 5. Distance between connected agents in miles.**

| Algorithm | Total Distance | | | AC Distance | | |
|---|---|---|---|---|---|---|
| | Mean | Median | Standard Deviation | Mean | Median | Standard Deviation |
| MPFY | 405.6 | 405.4 | 7.5 | 24.1 | 24.1 | 0.7 |
| MPAY | 140.7 | 140.7 | 0.6 | 19.7 | 19.8 | 0.3 |
| SDFY | 153.8 | 153.3 | 6.1 | 21.2 | 21.1 | 0.7 |
| SDAY | 130.5 | 130.3 | 0.8 | 20.6 | 20.6 | 0.4 |
| ACAY | 274.9 | 274.7 | 2.4 | 11.7 | 11.7 | 0.3 |

Figure 9 shows the profit per metric tonne by matching algorithm. Profit is the difference between expenses and revenue from tax incentives. The MPFY algorithm has the lowest profit per tonne, while the MPAY has the highest. The profits for SDFY, SDAY, and MPAY algorithms are close, which indicates that minimizing the total transportation distance is essential regardless of whether one seeks the immediate or the best future connection.

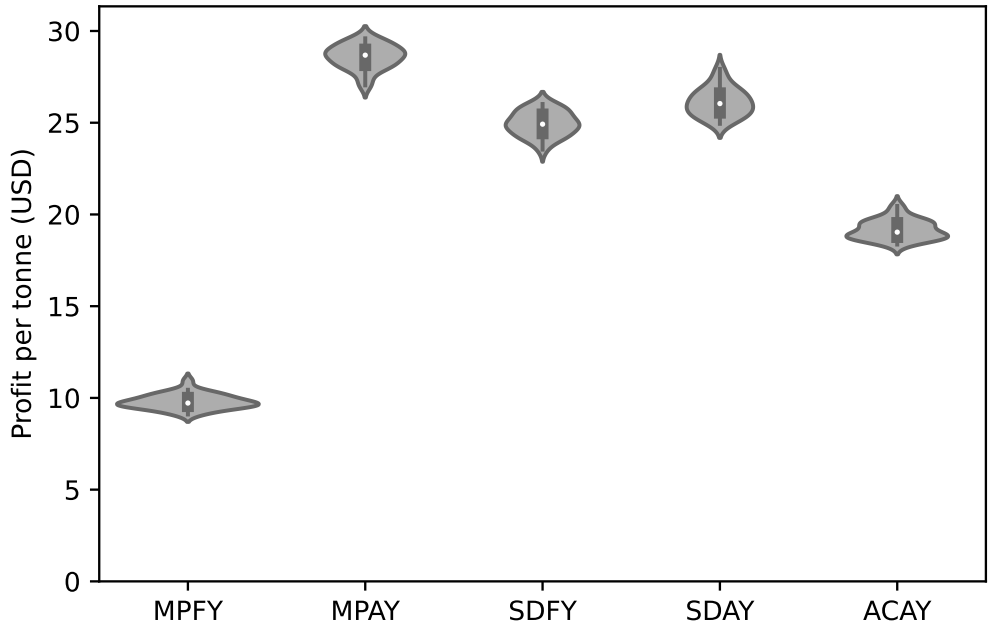

**Figure 9. Profit per tonne in USD by matching algorithm.**

Looking at the $CO_2$ captured by year, Figure 10 shows that the MPFY and SDFY algorithms capture more $CO_2$ earlier in the time horizon, while the MPAY and SDAY algorithms capture more later in the time horizon. A facility that starts capturing $CO_2$ in 2025 will capture it until 2037. In addition, if a connection is made during the simulation, it will always be connected from 2032 to 2037; hence, it will be the highest capture in those years. The capture amounts for these years are about 800 MT. Figure A.2 in the appendix shows the distribution of $CO_2$ transported by transportation mode for the algorithms.

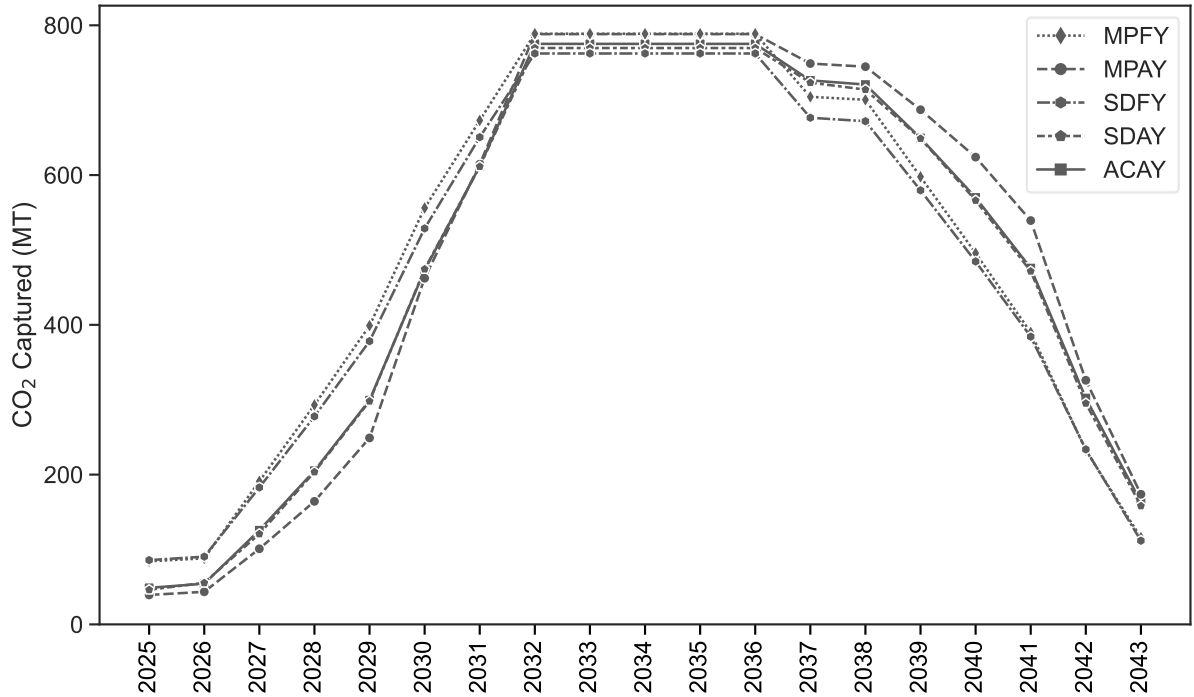

**Figure 10. CO$_2$ captured by year for each matching algorithm.**

Sensitivity analysis experiments were conducted for five primary cost parameters: capture, storage, pipeline transportation, rail transportation, and water transportation costs. These varied from 20% to 200%. Figure 11 presents the results for the MPAY algorithm, illustrated by the distribution of total CO$_2$ captured from 2025 to 2043. Changing the transport cost had little effect on the system; even varying them together had minimal impact. However, varying the capture cost (both increasing and decreasing) affected the amount captured. For example, if the capture costs are reduced by 40%, the total capture is nearly 22 GT, about 1.2 GT capture per annum on average. According to Fahs et al. (2023), capture costs represent the majority of cost reduction potential.

Another set of sensitivity experiments was conducted for the MPAY algorithm, which mandates capturing CO$_2$ from a baseline of 12 years up to 18 years, with the total revenue remaining constant. The total revenue generated from the connection duration between a supply agent and a demand agent was constant as the connection duration increased from 12, the baseline, up to 18 years, so the generated revenue from the tax credit of the first 12 years must also cover the operating cost of both agents and the additional transport cost for all years beyond year 12. There was a gradual decrease as the years increased until 16 when the total CO$_2$ captured dropped to near zero (Figure 12).

Figure 13 shows the different amounts of CO$_2$ captured with varying percent share of revenue with the supply agents. The capture increases gradually from 50% to 85%, then drops significantly with a higher percent share.

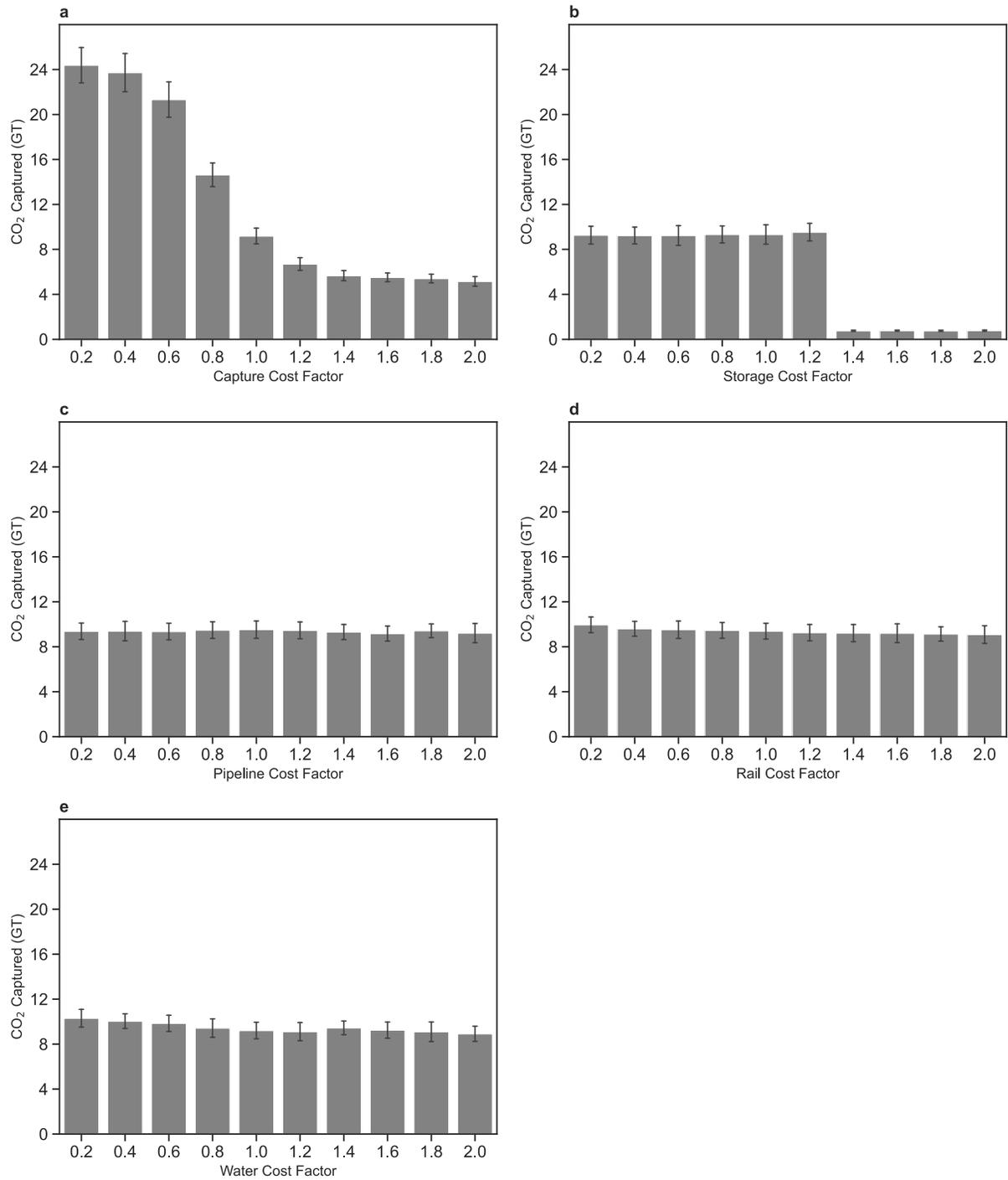

**Figure 11. Results for analysis of cost factors.**

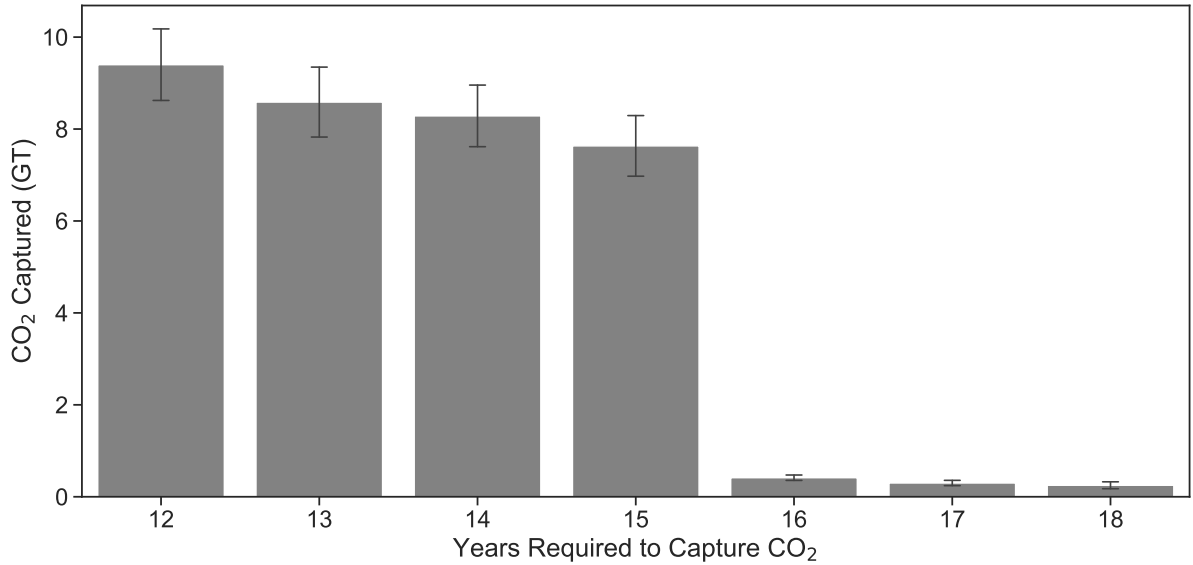

**Figure 12. Total CO$_2$ captured by the years mandated to capture CO$_2$ from the baseline of 12 years up to 18 years, with the total revenue remaining constant.**

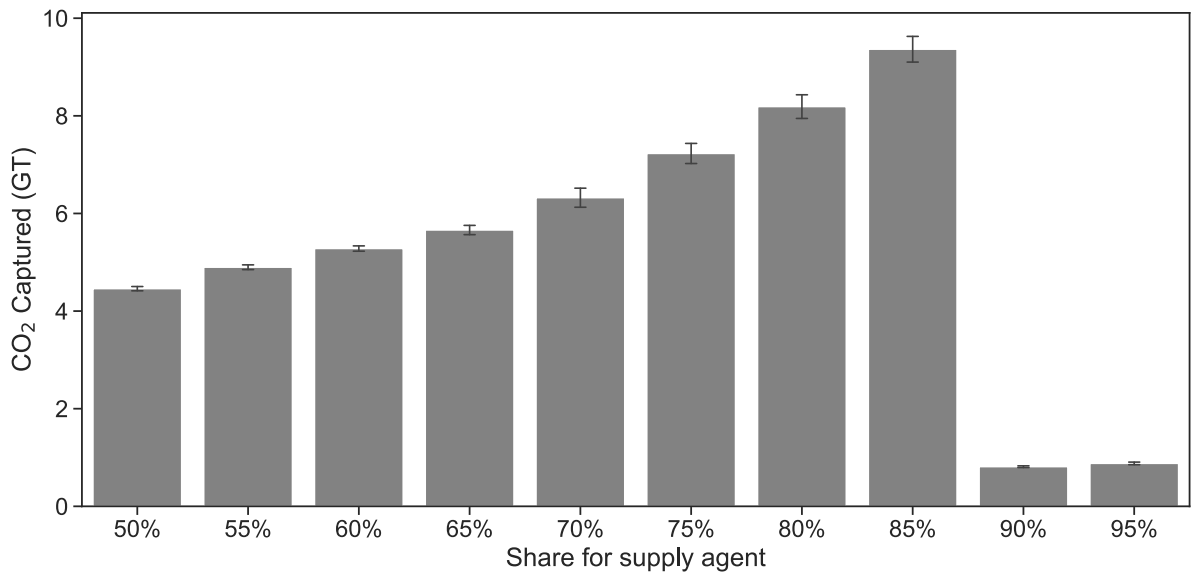

**Figure 13. Total CO$_2$ captured by varying revenue share amounts for the supply agents.**

## 5. Conclusion

This study developed and presented an agent-based model, CCUS-Agent, for multimodal transportation of CO$_2$ for carbon capture, utilization, and storage (CCUS). The tool was designed to maximize CO$_2$ capture while determining the potential transportation constraints the CCUS system might encounter over time. The tool incorporated real-world data regarding supply and demand agents, multimodal transportation networks, and 45Q tax credits in the United States. Five algorithms were tested to match supply and demand agents, emphasizing the market-based approach. The results from the case study indicated that over 9 billion tonnes of $CO_2$ can be captured from 2025 to 2043 while making sure the connections between supply and demand agents are profitable, i.e., the supply agent portion of revenue must cover the supply agent costs and all transportation costs, while the demand agent revenue portion must cover all demand agent costs. The total capture amount could be increased significantly with the reduction in capture costs. In future works, the authors would like to extend the tool to incorporate additional capabilities, e.g., model outputs with actual transportation routes and capacity for those routes.

## CRediT authorship contribution statement

**Majbah Uddin:** Conceptualization, Methodology, Formal analysis, Investigation, Writing - Original Draft. **Robin J Clark:** Conceptualization, Methodology, Software, Formal analysis, Investigation, Writing - Original Draft. **Michael R Hilliard:** Conceptualization, Methodology, Supervision, Funding acquisition, Writing - review & editing. **Josh A Thompson:** Conceptualization, Writing - review & editing. **Matthew H Langholtz:** Conceptualization, Writing - review & editing. **Erin G Webb:** Conceptualization, Writing - review & editing.

## Acknowledgment

This research is sponsored by the Laboratory Directed Research and Development Program of Oak Ridge National Laboratory, managed by UT-Battelle, LLC, for the U.S. Department of Energy.

# Appendix A

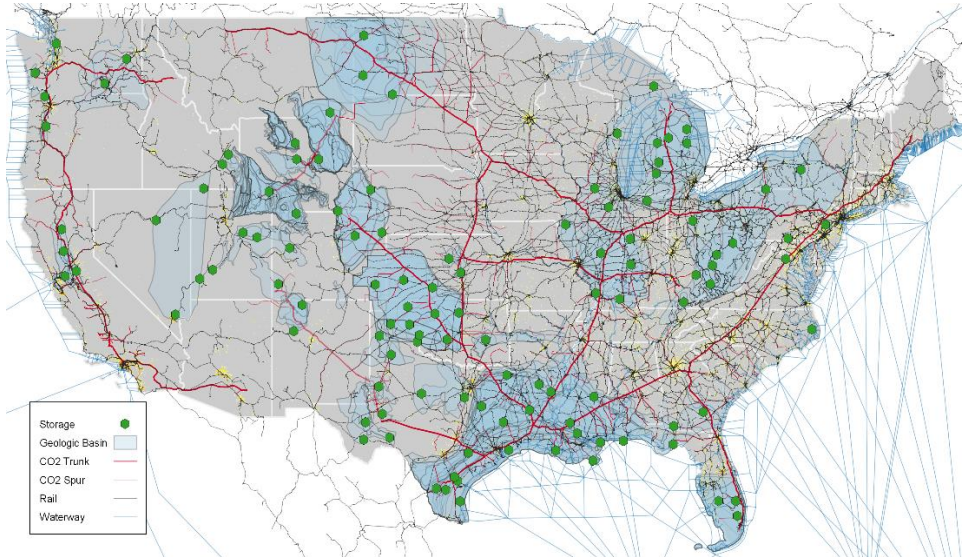

**Figure A.1. Identification of candidate storage location.**

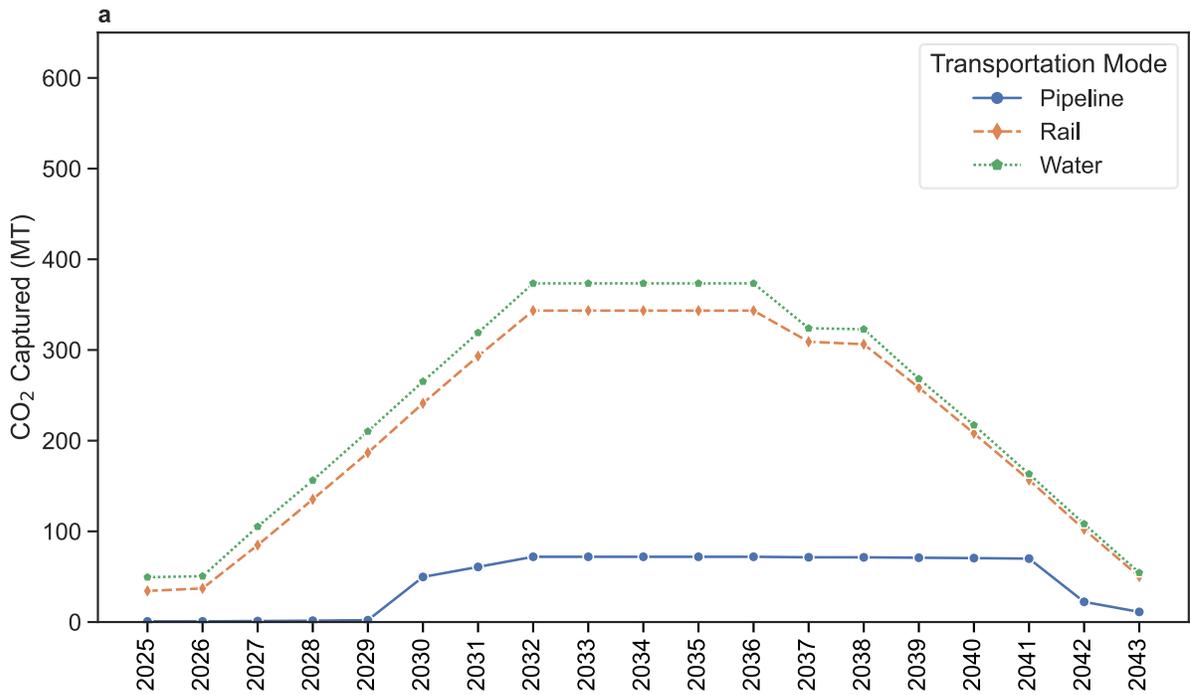

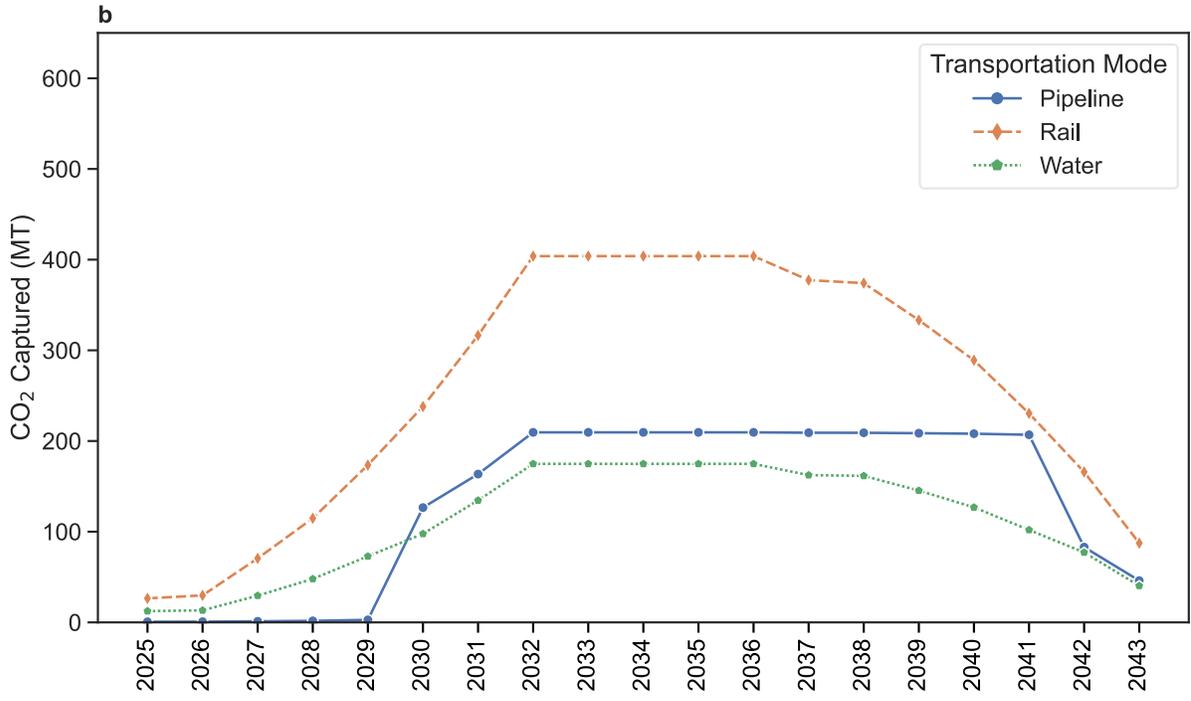

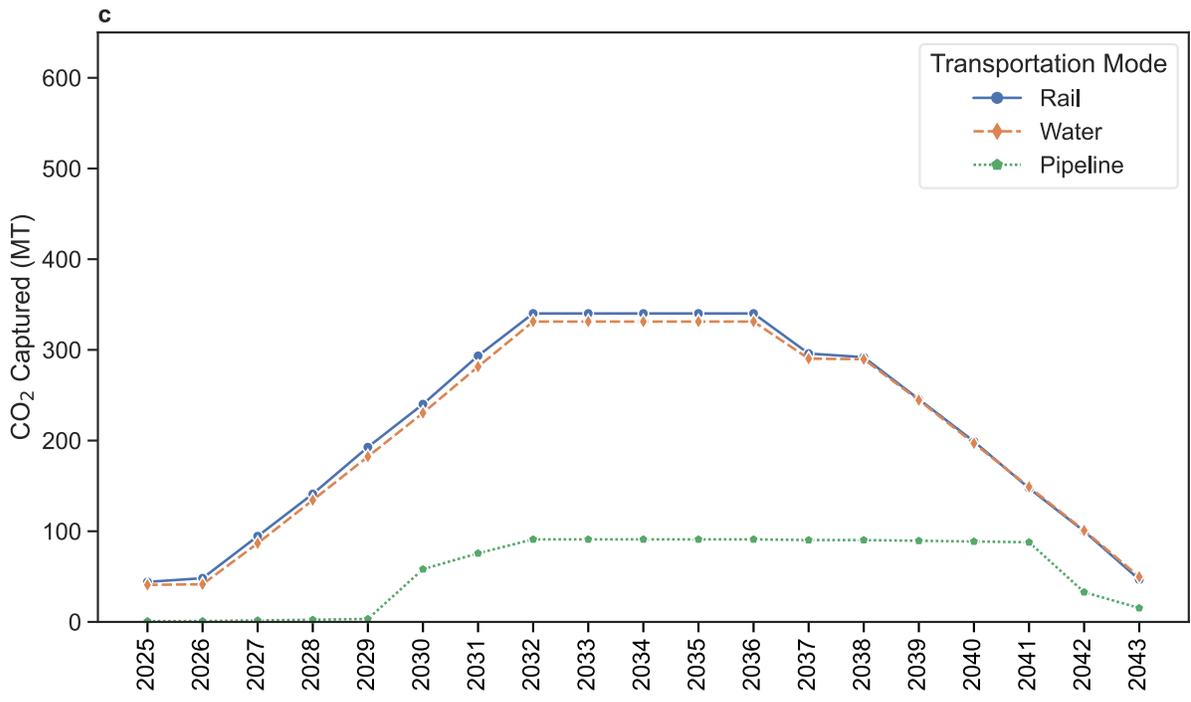

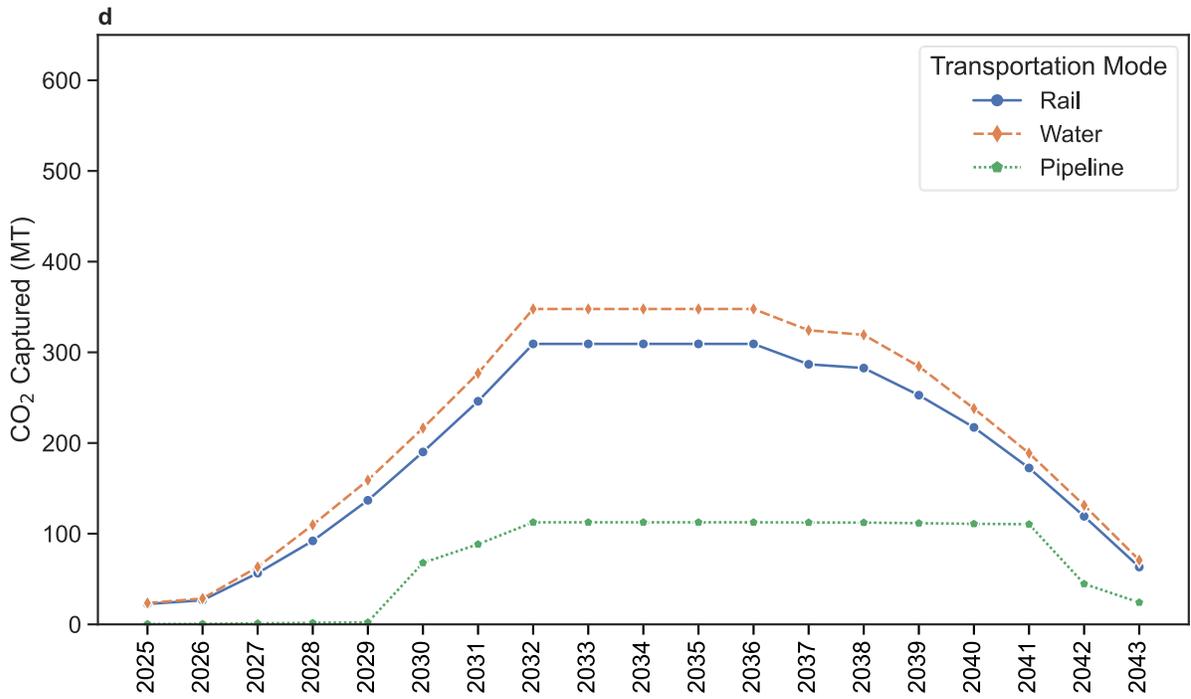

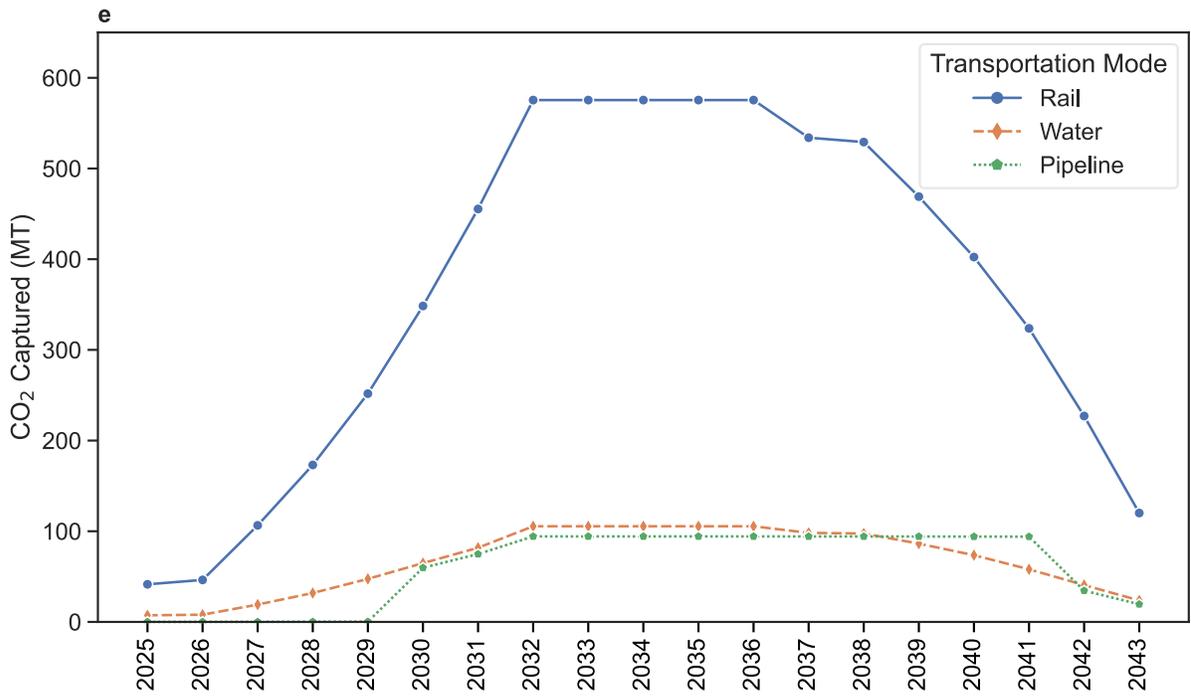

**Figure A.2. CO$_2$ transported by transportation mode by year: a) MPFY, b) MPAY, c) SDFY, d) SDAY, and e) ACAY.**